\documentstyle[twocolumn,aps,prl,array]{revtex}
\begin{document}
\draft
\title{Bell's theorem without inequalities and without
probabilities for two observers}
\author{Ad\'{a}n Cabello\thanks{Electronic address:
adan@cica.es}}
\address{Departamento de F\'{\i}sica Aplicada II,
Universidad de Sevilla, 41012 Sevilla, Spain}
\date{\today}
%First version: 4 August 2000
%This version: 8 February 2001
%Phys. Rev. Lett. {\bf 86}, 10, 1911-1914 (2001).
%quant-ph/0008085v4.
\maketitle
\begin{abstract}
A proof of Bell's theorem using two
maximally entangled states of two qubits is presented.
It exhibits a similar logical structure
to Hardy's argument of ``nonlocality without inequalities''.
However, it works for 100\% of the runs
of a certain experiment.
Therefore, it can also be viewed as a
Greenberger-Horne-Zeilinger-like proof
involving only two spacelike separated regions.
\end{abstract}
%Quantum optics

\narrowtext

%----------------------- INTRODUCTION -----------------------

Bell's theorem \cite{Bell64} states that one cannot in general reproduce
the results of quantum theory with a classical, deterministic local model.
Hardy's argument of ``nonlocality without inequalities'' \cite{Hardy93}
is considered ``the best version of Bell's theorem'' \cite{Mermin95}.
Curiously enough, while in the original proof of Bell's theorem
using inequalities the maximum discrepancy with local models
occurs for maximally entangled states \cite{Kar95}
(also known as ``Bell states'' \cite{BMR92}),
Hardy's proof is not valid for maximally entangled states.
It works only for entangled but non-maximally entangled states
(also known as ``Hardy states'' \cite{CN92}).
This curious feature has led to several attempts and
suggestions to develop a Hardy-like argument for
maximally entangled states of two qubits.
However, so far none of these proposals
has worked \cite{Cabello00a}.
In addition, Hardy's argument works only
for 9\% of the runs of a certain experiment.
On the other hand, the proof of Bell's theorem by Greenberger, Horne,
and Zeilinger (GHZ) \cite{GHZ89,Mermin90a,Mermin90b,GHSZ90}
works for 100\% of the runs, but requires
three observers (instead of two, as in Hardy's proof).

In this paper, I introduce a Hardy-like proof
which works for 100\% of the runs of an experiment.
It requires only two observers and is based on
maximally entangled states of two qubits. This proof
is more complicated than the original by Hardy, since
it involves two copies of the maximally entangled state
(instead of just one),
but it preserves the logical structure of the original argument.
The proof is inspired in
Hardy's argument \cite{Hardy93}
(see also \cite{Mermin95,CN92,Goldstein94,Mermin94})
and in a proof of the
Kochen-Specker theorem \cite{KS67} by Peres \cite{Peres90}
and Mermin \cite{Mermin90}
(see also \cite{Peres92,Mermin93,Peres93,Peres96}).

%----------------------- SCENARIO -----------------------

A qubit is a quantum two-level system like, for instance, the
spin state of a spin-$\frac{1}{2}$ particle.
Consider a source of two qubits in the singlet state
\begin{equation}
\left| \psi^- \right\rangle_{ij} =
{1 \over \sqrt{2}}
\left( {
\left| 01 \right\rangle_{ij} -
\left| 10 \right\rangle_{ij}
} \right),
\end{equation}
where $\left| 01 \right\rangle_{ij}=\left| 0 \right\rangle_{i}
\otimes \left| 1 \right\rangle_{j}$, and where
$\sigma_z \left| 0 \right\rangle =
\left| 0 \right\rangle$ and
$\sigma_z \left| 1 \right\rangle =
-\left| 1 \right\rangle$,
$\sigma_z$ and $\sigma_x$ being the Pauli spin matrices.
Suppose the source emits two singlets so that the initial state is
\begin{equation}
\left| \psi \right\rangle _{1234}=
\left| {\psi ^-} \right\rangle _{12} \otimes
\left| {\psi ^-} \right\rangle _{34}.
\label{twosinglets}
\end{equation}
The scenario for the proof is the following:
particle 1 moves away from particle 2, and particle
3 moves away from particle 4. At a given time, an observer,
Alice, has access to particles 1 and 3, while in a spacelike separated
region a second observer, Bob, has access to particles 2 and 4.

%----------------------- NOTATION -----------------------

I will use the following notation:
For particles 1 and 3 (Alice's particles)
$A_i = \sigma_{zi}$ and $a_i = \sigma_{xi}$;
for particles 2 and 4 (Bob's particles)
$B_j = \sigma_{zj}$ and $b_j = \sigma_{xj}$.
On the other hand, the product of two letters represents the
product of the corresponding operators; for instance,
$A_1 A_3 = \sigma_{z1} \otimes \sigma_{z3}$.
The results of the measurements of all these operators
can be either $-1$ or $+1$.

%----------------------- PROPERTIES -----------------------

Using this notation, the state (\ref{twosinglets}) has
the following properties:
\begin{eqnarray}
P_\psi \left( { A_1=B_2} \right) = 0,
\label{no1} \\
P_\psi \left( { a_1=b_2} \right) = 0,
\label{no2} \\
P_\psi \left( { A_3=B_4} \right) = 0,
\label{no3} \\
P_\psi \left( { a_3=b_4} \right) = 0,
\label{no4}
\end{eqnarray}
and
\begin{eqnarray}
P_\psi \left( {\left. B_2=B_4 \right| A_1 A_3=+1 } \right) = 1,
\label{Alice1} \\
P_\psi \left( {\left. b_2=b_4 \right| a_1 a_3=+1 } \right) = 1,
\label{Alice2} \\
P_\psi \left( {\left. A_1=a_3 \right| B_2 b_4=+1 } \right) = 1,
\label{Bob1} \\
P_\psi \left( {\left. a_1=-A_3 \right| b_2 B_4=-1 } \right) = 1,
\label{Bob2}
\end{eqnarray}
and
\begin{eqnarray}
P_\psi (& A_1 A_3=+1,&\,a_1 a_3=+1, \nonumber \\
 & B_2 b_4=+1,&\,b_2 B_4=-1) = \frac{1}{8},
\label{AliceandBob}
\end{eqnarray}
where, $P_\psi \left( { A_1=B_2} \right)$
is the probability of $A_1$ and $B_2$ giving the same result,
and $P_\psi \left( {\left. B_2=B_4 \right| A_1 A_3=+1 } \right)$
is the conditional probability of $B_2$ and $B_4$ having the same
result given that the result of $A_1 A_3$ is $+1$.
Properties (\ref{no1})-(\ref{AliceandBob}) will be proved later.
Now let us focus our attention on the logical argument of proving
Bell's theorem that can be deduced from them.

%----------------------- ARGUMENT -----------------------

Properties (\ref{no1})-(\ref{no4}) allow us to establish
that there exist ``elements of reality'', as defined by Einstein,
Podolsky, and Rosen (EPR) \cite{EPR35}, corresponding to
$A_1$, $B_2$, $a_1$, $b_2$, $A_3$, $B_4$,
$a_3$, and $b_4$. According to EPR,
{\em ``If, without in any way disturbing a system, we can predict
with certainty (i.e., with probability equal to unity)
the value of a physical quantity, then there exists an element of
physical reality corresponding
to this physical quantity''} \cite{EPR35}.
Property (\ref{no1}) tells us that if Alice (Bob) measures
$A_1$ on particle 1 ($B_2$ on particle 2), then she (he),
without in any way disturbing particle 2 (particle 1),
which is assumed to be in a distant spacelike separated region,
can predict with certainty the result of $B_2$ ($A_1$).
Therefore, there exists an element of reality
corresponding to $B_2$ ($A_1$). This implies that its value
was present before Alice's (Bob's) measurement, and
thus does not depend on Alice's (Bob's) choice of experiment.
Identical arguments based on properties (\ref{no2})-(\ref{no4})
lead us to establish that
there exist elements of reality corresponding to
$a_1$, $b_2$, $A_3$,
$B_4$, $a_3$, and $b_4$.

On the other hand, property (\ref{Alice1}) tells us that,
when Alice measures $A_1 A_3$ on particles $1$ and $3$,
and finds the result $+1$,
then she, without in any way disturbing particles 2 and 4,
which are assumed to be in a distant spacelike separated region,
can predict with certainty that if
Bob measures
$B_2$ and $B_4$, he will find the same result for both.
Since $B_2$ and $B_4$ are elements of reality, they
were determined before Alice's measurement and thus cannot
depend on Alice's choice of experiment.
Properties (\ref{Alice2})-(\ref{Bob2}) lead to similar arguments.

The four physical magnitudes $A_1 A_3$,
$a_1 a_3$, $B_2 b_4$, and $b_2 B_4$
are represented by commutative operators and
thus they can be jointly measured on the same system.
Indeed, according to property (\ref{AliceandBob}), the
four conditions appearing in (\ref{Alice1})-(\ref{Bob2})
(i.e., $A_1 A_3=+1$, $a_1 a_3=+1$, $B_2 b_4=+1$,
$b_2 B_4=-1$) can occur simultaneously (they occur in average
in $\frac{1}{8}$
of the runs of the experiment in which Alice measures $A_1 A_3$ and
$a_1 a_3$, while Bob measures $B_2 b_4$ and $b_2 B_4$).
Therefore, for those events
in which Alice finds $A_1 A_3=+1$ and
$a_1 a_3=+1$, and Bob finds $B_2 b_4=+1$ and $b_2 B_4=-1$,
the values of the elements of reality must satisfy
the following relations:
\begin{eqnarray}
v(B_2) & = & v(B_4),
\label{so1} \\
v(b_2) & = & v(b_4),
\label{so2} \\
v(A_1) & = & v(a_3),
\label{so3} \\
v(a_1) & = & -v(A_3).
\label{so4}
\end{eqnarray}
However, any assignment of values, either $-1$ or $+1$, that
satisfies (\ref{so1})-(\ref{so4}), would be in contradiction with one
of the properties (\ref{no1})-(\ref{no4}).
To prove this, let us suppose
that the four properties (\ref{no1})-(\ref{no4}) were satisfied.
Then, the values of the elements of reality would satisfy
the following relations:
\begin{eqnarray}
v(A_1) & = & -v(B_2),
\label{mo1} \\
v(a_1) & = & -v(b_2),
\label{mo2} \\
v(A_3) & = & -v(B_4),
\label{mo3} \\
v(a_3) & = & -v(b_4).
\label{mo4}
\end{eqnarray}
However, the eight Eqs.~(\ref{so1})-(\ref{mo4}) cannot be
satisfied simultaneously because
when we take the product of all of them, the result is
\begin{equation}
v(B_2) v(b_2) v(A_3) v(a_3) = -v(B_2) v(b_2) v(A_3) v(a_3).
\label{prod}
\end{equation}
Eq.~(\ref{prod}) has no solution because $B_2$, $b_2$,
$A_3$, and $a_3$ can take only the values $-1$ and
$+1$. We therefore conclude that
the predictions of quantum theory for these events cannot be
reproduced with any classical local model based on
EPR criterion of elements of reality.

%----------------------- ...AND "WITHOUT PROBABILITIES" -----------------------

As a close examination will reveal, a similar contradiction
can be found every time
the product of the results of (Alice's measurements)
$A_1 A_3$, $a_1 a_3$
(and of Bob's measurements), $B_2 b_4$,
and $b_2 B_4$ is $-1$, and there is no contradiction
if the product is $+1$.
In principle, there are 16 possible outcomes;
in eight of them, the product is $+1$,
and in the other eight, the product is $-1$.
It would therefore be interesting to calculate the probability
of occurrence of each of the 16 possible outcomes.
The results of these calculations
(which will be explained in more detail later) are in Table~I.
Surprisingly, as a scrutiny of Table~I shows,
only those events in which
the product of the results of $A_1 A_3$, $a_1 a_3$, $B_2 b_4$,
and $b_2 B_4$ is $-1$ have a nonzero probability to occur.
That is, {\em only those results that cannot be described
with local models occur}.
Therefore, according to the predictions of quantum mechanics,
all the events of the experiment in which Alice measures
$A_1 A_3$ and $a_1 a_3$, while Bob measures $B_2 b_4$ and
$b_2 B_4$, cannot be reproduced with any local model.

%----------------------- HARDY'S VS. THIS WORK -----------------------

This argument exhibits a similar logical structure to Hardy's.
Both require a series of experiments to establish that all the outcomes
that never happen (or always happen) are indeed never (or always) found.
In our case, these experiments are those to test quantum predictions
(\ref{no1})-(\ref{Bob2}).
Having done that, and having drawn inferences based on EPR criterion,
both arguments come to a crucial experiment.
In Hardy's crucial experiment a maximum
of 9\% of the runs \cite{Hardy93,Mermin95,Mermin94}
(or almost 50\% in its ``ladder'' version \cite{Hardy97,BBDH97})
yield results that are in contradiction with the EPR inferences.
In the proof presented here
the crucial experiment is the one whose results are in Table~I.
According to quantum mechanics,
all the runs of this experiment
yield results that are in contradiction with the EPR inferences.

%----------------------- GHZ VS. THIS WORK -----------------------

The fact that the contradiction occurs in all the runs resembles
the four-particle version
of the proof of Bell's theorem by GHZ \cite{GHZ89,GHSZ90}.
However, GHZ's proof is not based on
the state (\ref{twosinglets}) but in a ``GHZ state'' of four qubits.
The main difference is that while in the GHZ argument,
any EPR inference on one qubit requires a
measurement on the other three qubits; in our argument,
the EPR inferences on Alice's qubits require only
measurements on Bob's qubits,
that is, while GHZ need to consider four
(or three, in the three-particle version
\cite{Mermin90a,Mermin90b,GHSZ90})
different combinations of two spacelike separated regions,
the argument presented here needs only two spacelike
separated regions.

%----------------------- PROOF OF PROPERTIES 5 to 8 -----------------------

Let us now prove properties (\ref{Alice1})-(\ref{AliceandBob}).
For instance, property (\ref{Alice1}) is true if and only if
\begin{equation}
P_\psi \left( { A_1 A_3=+1} \right) > 0,
\label{gre}
\end{equation}
and
\begin{equation}
P_\psi \left( { A_1 A_3=+1,\,B_2=-B_4} \right) = 0.
\label{non}
\end{equation}
Condition (\ref{gre}) is fulfilled because
$P_\psi \left( { A_1 A_3=+1} \right)=\frac{1}{2}$.
Condition (\ref{non}) is also fulfilled
because the left hand side of Eq.~(\ref{non}) can be written as
\begin{eqnarray}
P_\psi & & \left( { A_1 A_3=+1,\,B_2=+1,\,B_4=-1} \right) +
\nonumber \\
P_\psi & & \left( { A_1 A_3=+1,\,B_2=-1,\,B_4=+1} \right) =
\nonumber \\
 & & P_\psi \left( { A_1=+1,\,A_3=+1,\,B_2=+1,\,B_4=-1} \right) +
\nonumber \\
 & & P_\psi \left( { A_1=-1,\,A_3=-1,\,B_2=+1,\,B_4=-1} \right) +
\nonumber \\
 & & P_\psi \left( { A_1=+1,\,A_3=+1,\,B_2=-1,\,B_4=+1} \right) +
\nonumber \\
 & & P_\psi \left( { A_1=-1,\,A_3=-1,\,B_2=-1,\,B_4=+1} \right),
\label{dem}
\end{eqnarray}
where the four probabilities appearing at the
right hand side of Eq.~(\ref{dem}) are zero,
due to the properties (\ref{no1}) and (\ref{no3}).
Similar arguments allow us to prove
properties (\ref{Alice2})-(\ref{Bob2}).

%----------------------- PROOF OF PROPERTY 9 -----------------------

To demonstrate property (\ref{AliceandBob}), it will be useful
to calculate the set of common vectors of $A_1 A_3$ and
$a_1 a_3$,
and the set of common vectors of $B_2 b_4$ and
$b_2 B_4$.
The common eigenvectors of $\sigma _z\otimes \sigma _z$ and
$\sigma _x\otimes \sigma _x$ are the Bell states
\begin{eqnarray}
\left| \phi^\pm \right\rangle & = &
{1 \over \sqrt{2}}
\left( {
\left| 00 \right\rangle \pm
\left| 11 \right\rangle
} \right),
\label{Phi} \\
\left| \psi^\pm \right\rangle & = &
{1 \over \sqrt{2}}
\left( {
\left| 01 \right\rangle \pm
\left| 10 \right\rangle
} \right).
\label{Psi}
\end{eqnarray}
These Bell states satisfy the following equations:
\begin{eqnarray}
\sigma _z\otimes \sigma _z\left| {\phi ^\pm } \right\rangle & = &
\left| {\phi ^\pm } \right\rangle,\\
\sigma _x\otimes \sigma _x\left| {\phi ^\pm } \right\rangle & = &
\pm\left| {\phi ^\pm } \right\rangle,\\
\sigma _z\otimes \sigma _z\left| {\psi ^\pm } \right\rangle & = &
-\left| {\psi ^\pm } \right\rangle,\\
\sigma _x\otimes \sigma _x\left| {\psi ^\pm } \right\rangle & = &
\pm\left| {\psi ^\pm } \right\rangle.
\end{eqnarray}
The common eigenvectors of $\sigma _z\otimes \sigma _x$ and
$\sigma _x\otimes \sigma _z$ are the Bell states
\begin{eqnarray}
\left| \chi^\pm \right\rangle & = &
{1 \over \sqrt{2}}
\left( {
\left| 0\bar{0} \right\rangle \pm
\left| 1\bar{1} \right\rangle
} \right),
\label{Chi} \\
\left| \omega^\pm \right\rangle & = &
{1 \over \sqrt{2}}
\left( {
\left| 1\bar{0} \right\rangle \pm
\left| 0\bar{1} \right\rangle
} \right),
\label{Omega}
\end{eqnarray}
where $\sigma_x \left| \bar{0} \right\rangle =
\left| \bar{0} \right\rangle$ and
$\sigma_x \left| \bar{1} \right\rangle =
-\left| \bar{1} \right\rangle$.
These Bell states satisfy the following equations:
\begin{eqnarray}
\sigma _z\otimes \sigma _x\left| {\chi ^\pm } \right\rangle & = &
\left| {\chi ^\pm } \right\rangle,\\
\sigma _x\otimes \sigma _z\left| {\chi ^\pm } \right\rangle & = &
\pm\left| {\chi ^\pm } \right\rangle,\\
\sigma _z\otimes \sigma _x\left| {\omega ^\pm } \right\rangle & = &
-\left| {\omega ^\pm } \right\rangle,\\
\sigma _x\otimes \sigma _z\left| {\omega ^\pm } \right\rangle & = &
\pm\left| {\omega ^\pm } \right\rangle.
\end{eqnarray}
Therefore, the probability appearing in
Eq.~(\ref{AliceandBob}) can be calculated as
the probability of finding the common eigenvector of
$\sigma_{z1} \otimes \sigma_{z3}$,
$\sigma_{x1} \otimes \sigma_{x3}$,
$\sigma_{z2} \otimes \sigma_{x4}$, and
$\sigma_{x2} \otimes \sigma_{z4}$,
with eigenvalues $+1$, $+1$, $+1$, and $-1$, respectively; that is
\begin{equation}
P=\left| {\left\langle {\phi ^+
\chi ^-} | {\psi } \right\rangle_{1324}} \right|^2,
\end{equation}
where $\left\langle {\phi ^+ \chi ^-} \right|  =
\left\langle {\phi ^+} \right| _{13} \otimes
\left\langle {\chi ^-} \right| _{24}$,
and $\left|{\psi } \right\rangle _{1324}$ is the state defined in
(\ref{twosinglets}), after permuting qubits 2 and 3.
To calculate $P$ it is useful to express the state
(\ref{twosinglets}) as
\begin{eqnarray}
\left| {\psi } \right\rangle_{1324} & = &
{1 \over 2 \sqrt{2}}
\left(\left| \phi ^+ \chi ^- \right\rangle +
\left| \phi ^+ \omega ^+ \right\rangle -
\left| \phi ^- \chi ^+ \right\rangle
\right. \nonumber \\ & & \left. +
\left| \phi ^- \omega ^- \right\rangle -
\left| \psi ^+ \chi ^+ \right\rangle -
\left| \psi ^+ \omega ^- \right\rangle
\right. \nonumber \\ & & \left. +
\left| \psi ^- \chi ^- \right\rangle -
\left| \psi ^- \omega ^+ \right\rangle\right).
\end{eqnarray}
Then is easy to obtain that
$P= \frac{1}{8}$.
A similar reasoning leads to each of the probabilities appearing in Table~I.

%----------------------- ALTERNATIVE PROOF OF PROPERTY 9 -----------------------

These probabilities can also be obtained by
realizing that a measurement of
$A_1 A_3$ and
$a_1 a_3$ is equivalent to a measurement
of the Bell operator whose eigenvectors are the Bell states
$\left\{ {\left| {\phi ^\pm } \right\rangle,\,
\left| {\psi ^\pm } \right\rangle } \right\}$
on particles 1 and 3. Such measurement induces, via
``entanglement swapping'' \cite{es},
that particles 2 and 4 collapse to the same Bell state
as particles 1 and 3.
For instance, Alice cannot obtain $A_1 A_3=+1$ and $a_1 a_3=+1$
(which is equivalent to obtaining $\left| {\phi ^+} \right\rangle_{13}$)
while Bob obtains $B_2 b_4=+1$ and $b_2 B_4=+1$
(which is equivalent to obtaining $\left| {\chi ^+} \right\rangle_{24}$),
because $\left| {\phi ^+} \right\rangle_{24}$
(the state induced by entanglement swapping) is orthogonal to
$\left| {\chi ^+} \right\rangle_{24}$ \cite{curiosity}.
%The relation between both bases of maximally entangled states is
%\begin{eqnarray}
%\left| {\chi ^+} \right\rangle & = & {1 \over \sqrt{2}} \left(
%\left| {\psi ^+} \right\rangle +\left| {\phi ^-} \right\rangle \right), \\
%\left| {\omega ^-} \right\rangle & = & {1 \over \sqrt{2}} \left(
%\left| {\psi ^+} \right\rangle -\left| {\phi ^-} \right\rangle \right), \\
%\left| {\chi ^-} \right\rangle & = & {1 \over \sqrt{2}} \left(
%\left| {\phi ^+} \right\rangle +\left| {\psi ^-} \right\rangle \right), \\
%\left| {\omega ^+} \right\rangle & = & {1 \over \sqrt{2}} \left(
%\left| {\phi ^+} \right\rangle -\left| {\psi ^-} \right\rangle
%\right),
%\end{eqnarray}

%----------------------- POSSIBLE EXPERIMENTS -----------------------

The proof of Bell's theorem
presented here can be translated into real experiments
in the same way as Hardy's proof can. An experiment to test
Hardy's proof consists in preparing a source of the required states
and performing several tests of the
required properties \cite{BBDH97,TBMM95,BDD97}.
The same strategy applies here.
The source must prepare four qubits in the state (\ref{twosinglets}), and
we must then test properties (\ref{no1})-(\ref{AliceandBob})
separately. From a theoretical point of view, the
only difficulty lies on testing property (\ref{AliceandBob}),
since it involves a joint measurement of $A_1 A_3$ and
$a_1 a_3$, and a joint measurement of $B_2 b_4$ and $b_2 B_4$.
However, as seen above, a joint measurement of $A_1 A_3$ and
$a_1 a_3$ ($B_2 b_4$ and $b_2 B_4$) is equivalent to measuring
a Bell operator whose eigenvectors are the Bell states
$\left\{ {\left| {\phi ^\pm } \right\rangle,\,
\left| {\psi ^\pm } \right\rangle } \right\}$
($\left\{ {\left| {\chi ^\pm } \right\rangle,\,
\left| {\omega ^\pm } \right\rangle } \right\}$).
Indeed, since
no complete discrimination
between the four Bell states is needed
to obtain an event which cannot be explained with local models
(it is enough to detect, for instance,
$\left| {\phi ^+ } \right\rangle_{13}$
and $\left| {\chi ^- } \right\rangle_{24}$),
then previous set-ups to distinguish between two of
the four Bell states in
the case of photons entangled in polarization
\cite{dc2,telep2}, can be used for this purpose.

If a complete discrimination between the four Bell states
$\left\{ {\left| {\phi ^\pm } \right\rangle,\,
\left| {\psi ^\pm } \right\rangle } \right\}$
($\left\{ {\left| {\chi ^\pm } \right\rangle,\,
\left| {\omega ^\pm } \right\rangle } \right\}$)
were possible
(for some proposals, see \cite{SEB99,DCD00,VFT00,PPB00}),
it would be interesting to experimentally verify
the predictions of quantum mechanics contained in Table~I.
That is, to verify whether, out of the 16 possible results
of the experiment in which Alice measures
$A_1 A_3$ and $a_1 a_3$, and
Bob measures $B_2 b_4$ and $b_2 B_4$
on the state given by Eq.~(\ref{twosinglets}), only
those eight that are unexplainable with local models have
a nonzero probability to occur.

%----------------------- ACKNOWLEDGEMENTS -----------------------

I thank David Mermin and
Asher Peres for their stimulating comments on
an earlier version and their
valuable suggestions for its improvement.

%----------------------- REFERENCES -----------------------

%\newpage

%----------------------- TABLE -----------------------

\begin{table}
\begin{center}
\begin{tabular}{ccccc}
\hline
\hline
$A_1 A_3$ & $a_1 a_3$ & $B_2 b_4$ & $b_2 B_4$ & Probability\\
\hline
$+1$ & $+1$ & $\pm1$ & $\pm1$ & $0$ \\
$+1$ & $+1$ & $\pm1$ & $\mp1$ & $0.125$ \\
$+1$ & $-1$ & $\pm1$ & $\pm1$ & $0.125$ \\
$+1$ & $-1$ & $\pm1$ & $\mp1$ & $0$ \\
$-1$ & $+1$ & $\pm1$ & $\pm1$ & $0.125$ \\
$-1$ & $+1$ & $\pm1$ & $\mp1$ & $0$ \\
$-1$ & $-1$ & $\pm1$ & $\pm1$ & $0$ \\
$-1$ & $-1$ & $\pm1$ & $\mp1$ & $0.125$ \\
\hline
\hline
\end{tabular}
\end{center}
%\vspace{0.2cm}
\noindent TABLE I.
{\small Probabilities of the 16 possible results
of the experiment in which Alice measures
$A_1 A_3$ and $a_1 a_3$ and
Bob measures $B_2 b_4$ and $b_2 B_4$
on the state given by Eq.~(\ref{twosinglets}).}
\end{table}
\end{document}